\documentclass[english,useAMS,usenatbib]{mn2e} 
\usepackage{lmodern}

\usepackage[T1]{fontenc}
\usepackage[utf8]{inputenc}
\usepackage{color}
\usepackage[usenames,dvipsnames]{xcolor}
\usepackage{amsmath}
\usepackage{amssymb}
\usepackage{graphicx}
\usepackage{esint}
\usepackage{placeins}
\usepackage[authoryear]{natbib}

\usepackage[draft]{hyperref}
\definecolor{darkgreen}{rgb}{0.1,0.6,0.7}
\hypersetup{colorlinks = true,urlcolor=blue,citecolor=Blue}

\makeatletter
\usepackage{lmodern}\usepackage{epstopdf}

\let\jnl@style=\rm
\def\ref@jnl#1{{\jnl@style#1}}

\def\aj{\ref@jnl{AJ}}                   
\def\actaa{\ref@jnl{Acta Astron.}}      
\def\araa{\ref@jnl{ARA\&A}}             
\def\apj{\ref@jnl{ApJ}}                 
\def\apjl{\ref@jnl{ApJ}}                
\def\apjs{\ref@jnl{ApJS}}               
\def\ao{\ref@jnl{Appl.~Opt.}}           
\def\apss{\ref@jnl{Ap\&SS}}             
\def\aap{\ref@jnl{A\&A}}                
\def\aapr{\ref@jnl{A\&A~Rev.}}          
\def\aaps{\ref@jnl{A\&AS}}              
\def\azh{\ref@jnl{AZh}}                 
\def\baas{\ref@jnl{BAAS}}               
\def\bac{\ref@jnl{Bull. astr. Inst. Czechosl.}}
\def\caa{\ref@jnl{Chinese Astron. Astrophys.}}
\def\cjaa{\ref@jnl{Chinese J. Astron. Astrophys.}}
\def\icarus{\ref@jnl{Icarus}}           
\def\jcap{\ref@jnl{J. Cosmology Astropart. Phys.}}
\def\jrasc{\ref@jnl{JRASC}}             
\def\memras{\ref@jnl{MmRAS}}            
\def\mnras{\ref@jnl{MNRAS}}             
\def\na{\ref@jnl{New A}}                
\def\nar{\ref@jnl{New A Rev.}}          
\def\pra{\ref@jnl{Phys.~Rev.~A}}        
\def\prb{\ref@jnl{Phys.~Rev.~B}}        
\def\prc{\ref@jnl{Phys.~Rev.~C}}        
\def\prd{\ref@jnl{Phys.~Rev.~D}}        
\def\pre{\ref@jnl{Phys.~Rev.~E}}        
\def\prl{\ref@jnl{Phys.~Rev.~Lett.}}    
\def\pasa{\ref@jnl{PASA}}               
\def\pasp{\ref@jnl{PASP}}               
\def\pasj{\ref@jnl{PASJ}}               
\def\rmxaa{\ref@jnl{Rev. Mexicana Astron. Astrofis.}}%
\def\qjras{\ref@jnl{QJRAS}}             
\def\skytel{\ref@jnl{S\&T}}             
\def\solphys{\ref@jnl{Sol.~Phys.}}      
\def\sovast{\ref@jnl{Soviet~Ast.}}      
\def\ssr{\ref@jnl{Space~Sci.~Rev.}}     
\def\zap{\ref@jnl{ZAp}}                 
\def\nat{\ref@jnl{Nature}}              
\def\iaucirc{\ref@jnl{IAU~Circ.}}       
\def\aplett{\ref@jnl{Astrophys.~Lett.}} 
\def\apspr{\ref@jnl{Astrophys.~Space~Phys.~Res.}}
\def\bain{\ref@jnl{Bull.~Astron.~Inst.~Netherlands}} 
\def\fcp{\ref@jnl{Fund.~Cosmic~Phys.}}  
\def\gca{\ref@jnl{Geochim.~Cosmochim.~Acta}}   
\def\grl{\ref@jnl{Geophys.~Res.~Lett.}} 
\def\jcp{\ref@jnl{J.~Chem.~Phys.}}      
\def\jgr{\ref@jnl{J.~Geophys.~Res.}}    
\def\jqsrt{\ref@jnl{J.~Quant.~Spec.~Radiat.~Transf.}}
\def\memsai{\ref@jnl{Mem.~Soc.~Astron.~Italiana}}
\def\nphysa{\ref@jnl{Nucl.~Phys.~A}}   
\def\physrep{\ref@jnl{Phys.~Rep.}}   
\def\physscr{\ref@jnl{Phys.~Scr}}   
\def\planss{\ref@jnl{Planet.~Space~Sci.}}   
\def\procspie{\ref@jnl{Proc.~SPIE}}   

\newcommand{\boS}{{\ensuremath{\sf{S}}}}

\newcommand{\mathd}{\ensuremath{\mathrm{d}}}

\newcommand{\calP}{\ensuremath{\mathcal{P}}}

\newcommand{\calH}{\ensuremath{\mathcal{H}}}

\title[The skewed weak lensing likelihood]{The skewed weak lensing likelihood: why biases arise, despite data and theory being sound.}

\author[Sellentin et al.]{Elena Sellentin$^{1}$, Catherine Heymans$^{2}$, Joachim Harnois-Déraps$^{2}$\\
$^{1}$Département de Physique Théorique, Université de Genève, 24 Quai Ernest-Ansermet, CH-1211 Genève, Switzerland\\
$^{2}$Scottish Universities Physics Alliance, Institute for Astronomy, University of Edinburgh, Royal Observatory,\\ Blackford Hill, Edinburgh, EH9 3HJ, UK.
}

\makeatother

\usepackage{babel}
\begin{document}
\setlength{\voffset}{-12mm}

\date{Accepted 300 BC. Received 800 AD; in original form 10.000 BC}

\maketitle
\pagerange{\pageref{firstpage}--\pageref{lastpage}} \pubyear{2016}

\label{firstpage} 
\begin{abstract}
We derive the essentials of the skewed weak lensing likelihood via a simple Hierarchical Model.  Our likelihood passes four objective and cosmology-independent tests which a standard Gaussian likelihood fails. We demonstrate that sound weak lensing analyses are naturally biased low, and this does not indicate any new physics such as deviations from $\Lambda$CDM.  Mathematically, the biases arise because noisy two-point functions follow skewed distributions.  This form of bias is already known from CMB analyses, where the low multipoles have asymmetric error bars. Weak lensing is more strongly affected by this asymmetry as galaxies form a discrete set of shear tracer particles, in contrast to a smooth shear field.  We demonstrate that the biases can be up to 30\% of the standard deviation per data point, dependent on the properties of the weak lensing survey.  Our likelihood provides a versatile framework with which to address this bias in future weak lensing analyses.

\end{abstract}


\section{Introduction}
Weak lensing has matured into a powerful cosmological observable, from which the cosmological parameters and the cosmological model can be inferred \citep{Joudaki16,KiDS,Troxel}. However, weak lensing is also known for being a systematics-driven observational technique. Often discussed sources of systematic uncertainties are Intrinsic Alignments \citep{Blazek,IABridle}, misestimates of photometric redshifts \citep{Redshift1,Redshift2}, and multiplicative and additive biases in the shape measurements \citep{Shape1,Shape2,Shape3,Fenech}. Here, we specialize on a further influential origin of systematics in weak lensing, namely the problem that the actual \emph{likelihood} with which to analyze the data has so far been insufficiently known. 

The fact that the weak lensing likelihood cannot be Gaussian was demonstrated in \citet{SHInsuff}, which revealed that the actual weak-lensing likelihood must be left-skewed. However, most current weak lensing analyses employ the Gaussian likelihood and commonly find lower values for the normalization of the power spectrum, $\sigma_8$, than the Planck analyses \citep{Planck2015}. This has lead to many publications questioning whether this is due to systematics, or incorrect physics \citep[e.g.][]{MacCrann}.

However, as the weak lensing likelihood is left-skewed in reality, this means it generates more data which fall below the mean, than above. This implies that any weak lensing observation is intrinsically very likely to yield a data vector whose weak-lensing amplitude is `surprisingly' low  and the surprise arises only because our scientific expectations are currently mostly trained by a Gaussian likelihood. A low lensing amplitude does then not indicate a flaw in the data, but rather in the expectations.

In this paper we derive the mathematical form of the skewed likelihood of weak lensing 2-point functions, and prove that it represents simulated data more faithfully than a Gaussian likelihood. The correct likelihood is a mandatory prerequisite to yield unbiased constraints on physical theories: without it, neither maximum-likelihood estimators for parameters, nor the goodness of fit or $p$-values, nor Deviance Information Criteria for sanity checks or model selection, nor Bayesian evidences can be computed without biases \citep{Trotta}. A sound, well-understood and high-quality likelihood is therefore essential for robust constraints on a physical theory. Here, we will hence take the principled approach and begin to carefully construct the weak lensing likelihood from a mathematical argumentation line which separately implements the different noise processes occurring in weak lensing. A series of similarly principled approaches have been taken when preparing for analyses of the cosmic microwave background (CMB) \citep{Hamimeche1, Hamimeche2, BJK1, BJK2, Hivon}. In this paper we extend these derivations for weak lensing applications. The upcoming sections will present the core of a novel modular likelihood, and we expect a series of sequential refinements in the future. In spirit, our work is closest to \citet[][henceforth AHJ1617]{Alsing2,Alsing1} where a Bayesian Hierarchical Model for weak lensing was constructed -- here we will however employ forward modeling of the actual estimator-based techniques that are widely employed by current weak lensing surveys; CFHTLenS, KiDS and DES \citep{Joudaki16,KiDS,Troxel}.

Central to understanding this paper is the insight that noise on the data already exists prior to any attempt of inferring parameters from the data. As such, there must exist a single unique and cosmology-independent distribution from which the actual weak lensing data are drawn, and it must be possible to compute this distribution. It must then also be possible to verify the distribution's level of realism without making reference to a cosmological model. In contrast, the quality of any likelihood should never be judged by whether it gives the `correct' answer for physical parameters: this would be akin to changing statistical elements in the analysis until they prefer the physics one wishes to find. The majority of this paper will hence be cosmology independent, including our verification test conducted in Sect.~\ref{Accuracy}. Keeping the statistical description of the data strictly separate from the cosmological parameter inference leads here to the central result that weak lensing data are biased low by \emph{themselves}.  When conducting parameter inference, this bias manifests itself as a $\sigma_8$ constraint that is lower than the input cosmology, but is neither a signal of new physics, nor an indication that shape measurements etc.~were biased. 

The route of deriving the skewed likelihood by mathematically following how noise in the large-scale structure and shear measurement combine into a total likelihood, has been taken because this specialization to the weak lensing error budget is more powerful than employing general results from the non-Gaussian literature. General results apply to non-weak-lensing-related situations as well and are therefore somewhat vague. This includes e.g.~the Edgeworth expansion, or copula likelihoods \citep{Sato,Simon,Hartl}. Furthermore, the mathematical derivation enables future sequential improvements, even though the likelihood here derived already performs better than a Gaussian likelihood (Sect.~\ref{Accuracy}).

To highlight the importance of first describing the \emph{data} correctly before \emph{explaining} them with a physical model, parameter inference will be conducted in a future paper. This paper concludes instead with a discussion of the arising biases and their implication for our physical inference, and whether or not these biases can be precluded.

\section{Mathematical Derivation of the Likelihood}
\subsection{The Sense of Bayesian Hierarchical Models}
Although cosmological data are random variables and therefore  drawn from probability \emph{densities}, most data analyses in cosmology are still cumulant-based. For weak lensing, the signal is the second cumulant of a sky map, namely a 2-point function. Of this 2-point function, the standard-approach computes its respective second cumulant, which is the covariance matrix. The covariance matrix thus includes contributions from the fourth cumulant of the original sky map, termed the parallelogram configuration of the 4-point function. The covariance matrix thereby drops all elements of the 4-point function which are not part of the parallelogram configuration, even though these elements are non-vanishing.

What this standard-approach ignores, is that a cumulant-based inference is only complete and self-contained, if the likelihood is Gaussian. There exists no other likelihood, which has a finite number of cumulants \citep{EW}. Bayesian Hierarchical Models therefore skip over a successive inclusion of ever higher cumulants, and directly update from a cumulant-based inference to a \emph{distribution}-based inference, whereby they can handle non-Gaussian data self-consistently.

Given that the weak lensing likelihood is meanwhile known to be non-Gaussian and skewed \citep{SHInsuff}, developing a Hierarchical Model to capture this non-Gaussianity has become inevitable. One Hierarchical Model for weak lensing has already been developed and applied for weak lensing data in harmonic space, see AHJ1617. As is typical for Bayesian Hierarchical Models, the work of AHJ1617 is however numerically rather complex, and it is therefore worth wondering to which extent the model of AHJ1617 can be simplified. Moreover, weak lensing is on the very first level measured on an object-basis in real space, namely the shear estimate per individual galaxy. Due to the complexity of the real space mask, it would then be preferable to analyze the data in real space rather than in harmonic space. In the following sections, we will show that a major simplification of the Bayesian Hierarchical Model from AHJ1617 and a transition to real space are indeed possible. We shall also demonstrate that our simplified Hierarchical Model succeeds in generating synthetic data which display the same statistical behaviour as simulated data.

\subsection{Weak lensing 2-point functions}
The cosmological signal whose likelihood we here derive are weak lensing 2-point functions. For a review of weak lensing, the reader is referred to \citet{BartelSchneid} and here only the essentials for the upcoming argumentation line are collected. We introduce the amplitude of shear power spectra as
\begin{equation}
 A = \frac{3}{2} \Omega_{\rm m} \left( \frac{H_0}{c} \right)^2,
\end{equation}
where $\Omega_{\rm m}$ is the matter density parameter, $c$ is the speed of light, and $H_0$ is the Hubble constant. The angular shear power spectra per redshift bin combination $\mu,\nu$ are,
\begin{equation}
 C^{\mu \nu} (\ell) = A^2 \int \mathd \chi \frac{q_\mu(\chi) q_\nu(\chi)}{f_{\rm K}^2(\chi)} P_{\rm m}\left(\frac{\ell + 0.5}{ f_{\rm K}(\chi)},\chi\right).
 \label{Cmunu}
\end{equation}
Here, $P_{\rm m}$ is the Fourier matter power spectrum, evaluated at $k$-mode $\ell/\chi$, at the redshift corresponding to comoving distance $\chi$. Throughout this paper, the validity of the Limber approximation shall be assumed, meaning Cartesian Fourier space and harmonic space are regarded on an equal footing. For weak lensing, this is an excellent approximation and its accuracy is discussed in \citet{Limber0,Limber1,Limber2,Limber3,Limber4}. 
Tomography is enforced by splitting the galaxy populations into bins labeled by Greek indices, which leads to the lensing kernels,
\begin{equation}
 q_\mu(\chi) = \frac{f_{\rm K}(\chi)}{a(\chi)} \int_\chi^\infty \mathd \chi' n_\mu(\chi') \frac{f_{\rm K}(\chi'-\chi)}{f_{\rm K}(\chi')},
 \label{q}
\end{equation}
where $n_\mu(\chi)$ is the comoving distribution of galaxies in redshift bin $\mu$.
Given the shear power spectrum $C^{\mu\nu}_\ell$, a multitude of real-space correlation functions can be computed, by transforming via different filter functions to real space. In general, this transformation can be written as
\begin{equation}
\xi_{\rm F}(\theta) = \frac{1}{2\pi} \int \mathd \ell\ \ell F(\ell\theta) C^{\mu\nu}(\ell),
\label{filter}
\end{equation}
where $\xi_{\rm F}(\theta) $ is the real-space correlation function and $F(\ell\theta)$ is the filter that translates from harmonic space to real space. For a recent overview of the different filters commonly used in weak lensing, see \citet{Kilb}.

The most commonly used filters are the Bessel functions $J_0(\ell\theta)$ and $J_4(\ell\theta)$ which give rise to the 2-point correlation functions which can easily be computed from a galaxy catalogue, namely
\begin{equation}
 \xi_+(\theta) = \frac{1}{2\pi} \int \mathd \ell\ \ell J_0(\ell\theta) C^{\mu\nu}(\ell),
\end{equation}
and
\begin{equation}
 \xi_-(\theta) = \frac{1}{2\pi} \int \mathd \ell\ \ell J_4(\ell\theta) C^{\mu\nu}(\ell).
\end{equation}
In the following, the discussion will be mainly focused on $\xi_+(\theta)$, where this choice is representative of the other correlation functions as well: due to the filter $F(\ell \theta)$ in Eq.~(\ref{filter}) being linear, the upcoming statistical derivations carry through for any such filter, and focusing the discussion on $\xi_+$ is then not a limitation of the generality. This also holds true for further linear filtering to optimally compress the data, as done in \citet{Cos17,Moped}.

\subsection{A simple Hierarchical Model}
Given the need for a non-Gaussian likelihood, and the potential of Bayesian Hierarchical Models to provide principled and realistic solutions for this, one quickly arrives at the wish to simplify and verify the Bayesian framework for weak lensing as presented in AHJ1617. We here achieve this as follows.

One of the computationally most demanding steps in AHJ1617 arises because this model samples from a sky map on an intermediate step. It can thereby correctly include even highly complicated survey masks, encompassing the survey footprint and stellar and satellite masks alike. The parameter inference is then carried out in another step where the model transforms to harmonic space, where the shear power spectrum is compared to a theoretical prediction. To also include additive shape noise, the model of AHJ1617 employs numerically demanding Wiener filtering of the sky map in a third step. This computational hurdle has there been overcome via messenger fields \citep{Messenger1, Messenger2} and the overwhelming dimensionality of the data space to be sampled has been addressed with Gibbs and Hamilton Monte Carlo Sampling.

Although this framework has been proven to be numerically viable, the following observation indicates that a substantial simplification should be possible: the weak lensing 2-point functions $\xi_+,\xi_-$ are readily measured in real-space via the estimator
\begin{equation}
 \hat{\xi}^{\mu\nu}_\pm (\theta) = \frac{\sum_{ab} w_a w_b \left[ \epsilon_t^\mu(\vec{x}_a) \epsilon_t^\nu(\vec{x}_b) \pm \epsilon_\times^\mu(\vec{x}_a) \epsilon_\times^\nu(\vec{x}_b)\right]  }{\sum_{ab}w_aw_b},
\end{equation}
where $\epsilon = \epsilon_s + \gamma$, for weak shears when the measured ellipticities are the sum of source ellipticities $\epsilon_s$ and shear $\gamma$. The indices $\mu,\nu$ denote the redshift bins. This estimator counts galaxy pairs with a certain angular distance: The sum over $a,b$ runs over all galaxy pairs for which the angular separation $|\vec{x}_a - \vec{x}_b|$ falls into an interval $\theta \pm \Delta \theta$. As such, it reacts to the total number of pairs found, but as long as the noise is isotropic and homogeneous throughout the survey, and as long as the 2-point correlation function truly only depends on the distance between galaxies, this estimator is fairly blind with respect to the precise geometry of a mask, especially on angular separations much smaller than the survey footprint. But if the precise geometry of the survey volume has only a minor impact, then sampling from a sky map might not be necessary. Instead the 2-point correlation functions can be used as summary statistics.

Our final weak lensing data set originates from a map of the shear field on the sky. We denote the shear field as $S(\vartheta, \phi, \mu)$ where $\vartheta,\phi$ are the celestial angles, and $\mu$ labels the redshift bin, the field can be decomposed into spherical harmonics via
\begin{equation}
 S(\vartheta,\phi,\mu) = \sum_{\ell,m} a_{\ell m}^\mu Y_{\ell m}(\vartheta,\phi),
\end{equation}
where $Y_{\ell m}$ are the spherical harmonics and tomography is achieved by splitting the population of source galaxies into different distributions $n_\mu(z)$, see Eq.~(\ref{q}). Assuming statistical isotropy, and that the $a_{\ell m}^\mu$ are drawn from a Gaussian distribution with vanishing mean, the complete information contained in this data set is preserved when marginalizing out the individual modes, and transforming onto their 2-point correlation function instead. As the shear field is mildly non-Gaussian, this compression into a 2-point function will lose some information which could be assessed by sampling from the full map. This loss of information is however not necessarily related to a loss of valuable physical information: as long as precise theoretical predictions are available for 2-point functions only, cosmological parameter inference is not disadvantaged by here transforming into 2-point functions.

On the full sky, a realization of the thus arising angular power spectrum is given by
\begin{equation}
 \hat{C}_\ell = \frac{1}{2\ell +1} \sum_{m = -\ell}^{\ell} |a_{\ell m}|^2.
 \label{Cell}
\end{equation}
The realized power spectrum $\hat{C}_\ell$ does not follow a Gaussian distribution, since it sums up quadratic combinations of Gaussianly distributed variables. Squaring is a non-linear operation, and consequently Gaussianity is lost and $\hat{C}_\ell$ follows a left-skewed Gamma-distribution instead. Only for high $\ell$, where the sum in Eq.~(\ref{Cell}) runs over sufficiently many $m$-modes, does the Central Limit Theorem overwhelm the non-linearity of the quadratic estimator.

Inferring parameters from the shear power spectrum is observationally however not ideal. It is much easier to predict from a noisy full-sky power spectrum the corresponding real-space correlation function, than it is to infer a robust estimator of pseudo-$C_\ell$ by transforming a noisy real-space correlation function. This is because applying a mask is easier than deconvolving a mask, and because shape noise is trivially additive in real space, but not in harmonic space \citep{Alsing1,Alsing2,Hamimeche1,Hamimeche2}.

Returning to Eqs.~(\ref{Cmunu},\ref{q},\ref{filter}) we see however that the angular filter functions Eq.~(\ref{filter}) are noise-free weights. As the statistical distribution of the full-sky $\hat{C}_\ell$ is known, the noise of the $\hat{C}_\ell$ can be fed through to real-space, and there shape-noise can be added.

This forward modeling has a further advantage: As power spectra measure the variance in harmonic or Fourier space, they must be positive definite. In contrast, correlation functions can take negative values because they measure the excess probability of finding pairs with distance $r$. This is best known from the galaxy correlation function $\xi_{gg}$, where the probability to find pairs within a distance of $r$ is
\begin{equation}
 \langle n_{\rm pairs}\rangle = \bar{n}^2 [1+\xi_{gg}(r)] \mathd V_1 \mathd V_2,
\end{equation}
where $V_i$ are two infinitesimal volumes and $\bar{n}$ is the average spatial galaxy number density. This excess probability can be larger or lower than the average $\bar{n}^2$, indicating positive or negative correlation. The possible negativity implies immediately that correlation functions cannot be Gamma-distributed, since Gamma distributions generate positive semi-definite variables only.

Therefore, we exploit the knowledge of $\hat{C}_\ell$ being positive and Gamma-distributed, and feed this through to real-space by applying filter functions. This forward modeling naturally enables noisy negative values of the correlation function because the filter functions between harmonic and real space (here the Bessel functions) oscillate and admit negative values\footnote{To avoid confusion, we explicitly note that we are here \emph{not} discussing the problem of negative likelihoods, as they appear in an Edgeworth expansion. Our likelihood is positive definite but enables the necessary negative values of the data.}.

Accordingly, we start from the Gamma distribution of the noisy $\hat{C}_\ell$, given by \citet{MardiaKentBibby,AndersonTW,GuptaNagar,Hamimeche1,Hamimeche2} as
\begin{equation}
 \calP(\hat{C}_\ell | C_\ell) \propto \frac{ \hat{C}_\ell^{ \frac{\nu -2}{2}   } }{C_\ell^{\frac{\nu}{2}}} \exp\left( -\frac{\nu}{2} \frac{\hat{C}_\ell}{C_\ell} \right).
 \label{celldist}
\end{equation}
The scalar $\nu$ is called the degrees of freedom, and counts the number of modes averaged over. It thus depends on $\ell$ and for a full-sky observation, one would have $\nu = 2\ell +1$. 

The distribution Eq.~(\ref{celldist}) is a special type of a Gamma function.
If $x_i$ are generic Gamma distributed variables\footnote{We adopt the usual statistical convention, denoting `drawn from' as `$\sim$'.}, with
\begin{equation}
x_i \sim {\rm Gamma}(a_i, b) \, ,
\end{equation}
where $a$ is the shape parameter ($\nu/2$ in our case) and $b$ the scale parameter ($2C_\ell/\nu$ in our case) of the Gamma distribution, then the sum over such Gamma distributed samples follows 
\begin{equation}
 \left( \sum_{i = 1}^N x_i \right) \sim {\rm Gamma}\left[ \left(\sum_{i=1}^N a_i\right),b \right].
\end{equation}
Furthermore, a scalar multiple of a Gamma distributed variable is distributed according to 
\begin{equation}
{\rm if\ } x \sim {\rm Gamma(a,b)} \Rightarrow cx \sim {\rm Gamma}(a,cb).
\end{equation}
From these two properties, we see that there exists no analytical solution if we wish to add up multiple Gamma distributed $\hat{C}_\ell$ that are all drawn from different $\nu$, and that each are weighted with different prefactors arising from the filter function, $c = F(\ell\theta)$. Fortunately, sampling from a Gamma distribution is straight forward, such that a numerical implementation of this process is easy and has short code-execution times.

Drawing realizations $\hat{C}_\ell$ of the shear power spectrum includes randomness from the large-scale structure into the total weak lensing likelihood. A further dominant uncertainty is shape noise, generated because real galaxies are not spherical, but come with intrinsic source ellipticities, $\epsilon_{\rm s}$. The distribution of $\epsilon_{\rm s}$ is traditionally approximated by a Gaussian\footnote{For the purposes of this paper we adopt Gaussian- distributed shape noise, noting that in contrast to a Gaussian likelihood our analysis framework can readily include more complex distributions, such as the observed galaxy ellipticity distributions from, for example, \citet{melchior_viola2012,Shape3,Fenech}, that are not well approximated by Gaussian distributions.} with standard deviation $\sigma_{\epsilon_i} = 0.29$ per ellipticity component $(i=1,2)$ \citep[see for example][]{KiDS}.  These two dominant sources of scatter in weak lensing  combine into the likelihood
\begin{equation}
\begin{aligned}
\forall \ell&: \hat{C}_\ell \sim {\rm Gamma}\left[\frac{\nu(\ell)}{2}, \frac{2C_\ell}{\nu(\ell)}\right],\\
\forall \theta, \forall F&: \hat{\xi}_F(\theta) \sim \sum_\ell \frac{ \ell F(\ell \theta)}{2\pi}\hat{C}_\ell,\\
\hat{\xi}_F(\theta) \rightarrow  \hat{\xi}_F&(\theta) + s(\theta), {\rm \ with \ } s(\theta) \sim \mathcal{G}\left(0,{\sf{C}}_s\right),
\label{Like}
\end{aligned}
\end{equation}
where $\mathcal{G}$ is the Gaussian distribution and the shape noise $s(\theta)$ has covariance
\begin{equation}
{\sf{C}}_s = \frac{\sigma_\epsilon^2}{A_{\rm sky} \bar{n}^{2} 2\pi \theta \Delta\theta} \, ,
\end{equation}
where $\sigma_\epsilon^2 = \sigma_{\epsilon_1}^2 + \sigma_{\epsilon_2}^2$.
This covariance simply suppresses the scalar shape noise $\sigma_\epsilon$ per galaxy by the total number of galaxies averaged over in the angular bin $\Delta \theta$ around $\theta$ \citep{schneider/etal:2002,Covs}.

Eq.~(\ref{Like}) is the central result of this paper: it generates the distribution, $\mathcal{D}$, of weak lensing data, $\hat{\xi}_F(\theta) \sim \mathcal{D}_F(\hat{C}_\ell, \sigma^2_\epsilon, \bar{n},A_{\rm sky})$, which depends on the noise of the shear $\hat{C}_\ell$, the filter function $F$, shape noise $\sigma^2_\epsilon$, and survey area $A_{\rm sky}$ and the survey's galaxy density $\bar{n}$.

Eq.~(\ref{Like}) describes the noise on weak lensing data, independently of any cosmological model. Upon availability of a data vector and a cosmological theory, it therefore induces the weak lensing \emph{likelihood}
\begin{equation}
\mathcal{L}(\hat{\xi}_F^{\rm obs}|\boldsymbol{p}), 
\end{equation}
where $\hat{\xi}_F^{\rm obs}$ are the observed data, and $\boldsymbol{p}$ are the parameters to be inferred via maximum-likelihood estimation. If the order of the conditionality statement is reversed by priors $\pi$
\begin{equation}
\mathcal{L}(\boldsymbol{p}| \hat{\xi}_F^{\rm obs}) = \frac{\mathcal{L}(\hat{\xi}_F^{\rm obs}|\boldsymbol{p}) \pi(\boldsymbol{p})}{\pi(\hat{\xi}_F^{\rm obs})},
\end{equation}
then maximum-a-posteriori inference can be carried out.

The remaining task at hand, is now to work out the degrees of freedom, $\nu$. If a smooth field could be observed on the full sky, then the degrees of freedom would simply be $\nu = 2\ell +1$. As realistic surveys only cover a fraction of the sky, this has to reduce the degrees of freedom whereby statistical scatter is enhanced. Furthermore, galaxies are discrete tracer particles, rather than a smooth random field, and this must lead to a further reduction in the degrees of freedom.

It is well known that a survey which covers the fraction $f_{\rm sky}$ of the full sphere, can only support a finite number of correlated $\ell$ modes, and the degrees of freedom can then be approximated by $\nu \approx f_{\rm sky}(2\ell +1)$ \citep{Hamimeche1,Hamimeche2}.
The loss of statistical precision due to galaxies being discrete tracer particles can be estimated by enforcing a pixelization of the survey, such that sufficiently many galaxies in one pixel can jointly mimic a smooth but pixelized field.

We therefore imagine pixels on the sky with side length $\theta$. In the flat sky approximation the pixel area will be $A_{\rm pix} \approx \pi^2/\ell_{\rm pix}^2$. If the pixel is supposed to represent a smooth shear field, then it must contain a certain number $N_{\rm gal}$ of galaxies, that are averaged over. The higher the galaxy density of the survey, the smaller these pixels can be. If the survey has a number density of $\bar{n}$ galaxies per area, then a pixel that contains $N_{\rm gal}$ galaxies must have size $A_{\rm pix} = N_{\rm gal}/\bar{n}$, from which we identify,
\begin{equation}
 \ell_{\rm pix} \approx \sqrt{\frac{\pi^2\bar{n}} {N_{\rm gal}}}.
\end{equation}
The degrees of freedom on the masked and pixelated sky are then,
\begin{equation}
 \nu \approx f_{\rm sky} \frac{(2 \ell +1)}{\ell_{\rm pix}}.
\end{equation}
The degrees of freedom derived from this argumentation line have the right order of magnitude, but in the end it has to be underlined that this is an approximation. However, a refinement which would also allow to include soft variations induced by the harmonic transform of a survey mask, would be to include a factor $g_{\rm eff}(\ell)$ into the degrees of freedom.
 For simple masks, this factor can be computed analytically, but given the complications with survey boundaries and stellar masks, the factor $g_{\rm eff}$ is measured more easily from 200-300 simulations or from a Bayesian Hierarchical Model, \citep[see also the discussion][]{Hivon}. This leads to the total degrees of freedom being
\begin{equation}
 \nu \approx f_{\rm sky} \frac{(2 \ell +1)}{\ell_{\rm pix}} g_{\rm eff}(\ell).
 \label{eqn:geff}
\end{equation}

For the scope of this paper, we measure $g_{\rm eff}$ from the 930 SLICS weak lensing simulations \citep{SLICS}. We first determine the marginal densities of each data point $\xi_+(\theta_i)$, given by the histogram $\calH^i$ of the 930 SLICS samples.  These marginal distributions are readily predicted from our likelihood, given a value for $g_{\rm eff}(\ell)$. We hence determine our likelihood for different values of $g_{\rm eff}(\ell)$, each time drawing 930 samples per data point, and distributing them onto histograms $H^i$. The value for $g_{\rm eff}(\ell)$ that reproduces the simulations best, is then the value that minimizes the distance between the simulated histograms $\calH^i$ and the predicted histograms $H^i$. A stable metric to compute the distance between histograms is given by the $L_1$-norm and we hence minimize the total error
\begin{equation}
E_{\rm tot} = \sum_{i = 1} E_i,
\end{equation}
where the discrepancy between the histograms of each marginal distribution is
\begin{equation}
E^i = \frac{1}{B} \sum_{b = 1}^B | \calH^i_b - H_b^i  |,
 \label{Sdiv}
\end{equation}
where the subscript $b$ runs over the number of bins $B$ in each histogram.

We find that an $\ell$-independent $g_{\rm eff}$ is sufficient within the precision enabled by the simulations.  As a best-fitting value, we found $g_{\rm eff} \approx 2.29$, where the uncertainty on $g_{\rm eff}$ is caused by the limited number of 930 simulations. Fortunately, the shape and the amplitude of the likelihood is relatively stable to changes in $g_{\rm eff}$ of up to 40 percent. For the scope of this work, $g_{\rm eff}$ is therefore sufficiently well determined.  For future research we will target a deeper mathematical understanding of $g_{\rm eff}$ and thereby become independent of simulations.

\begin{figure*}
\includegraphics[width=\textwidth]{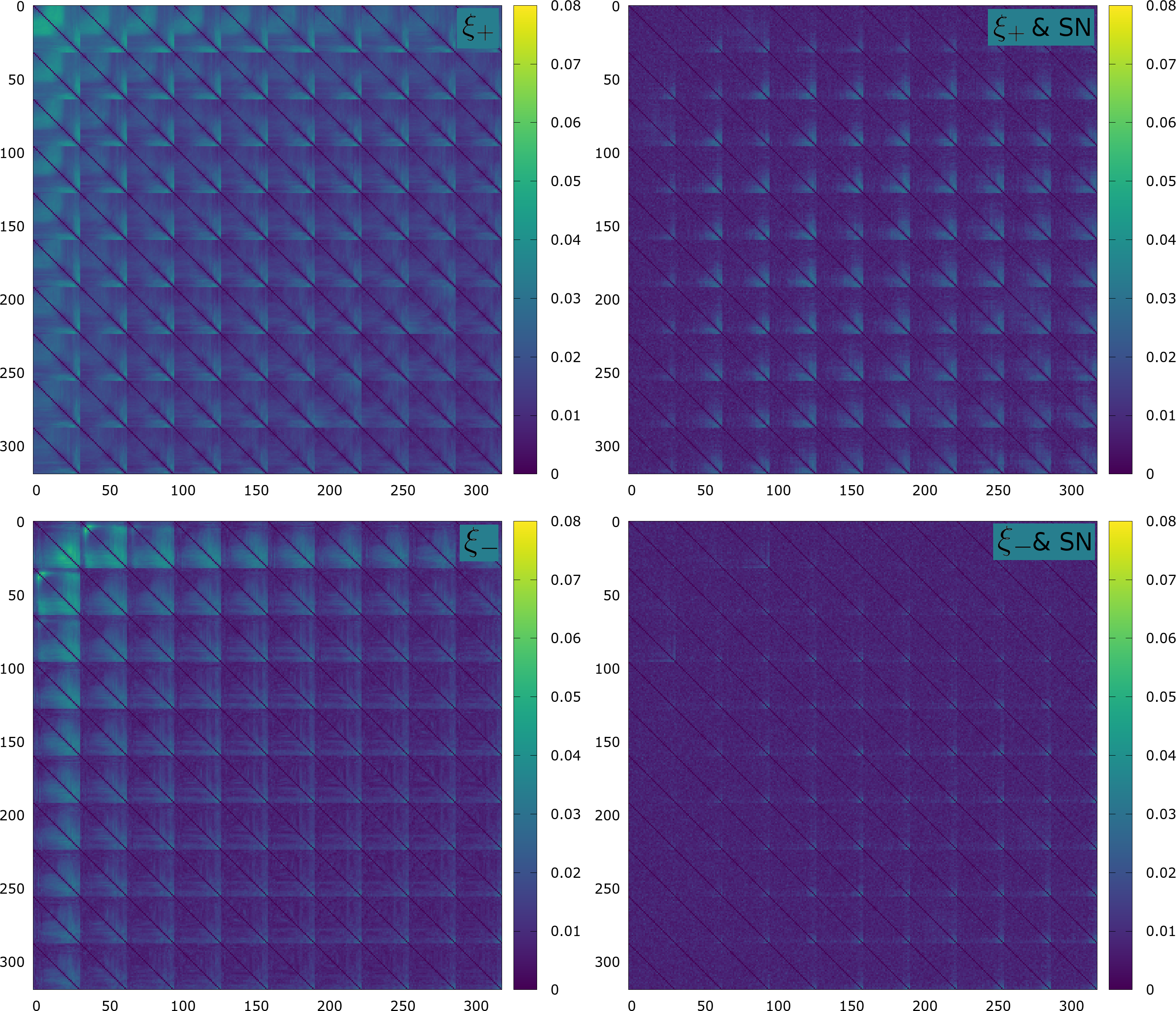} 
\caption{Trans-covariance matrices $\boS^+$ (Eq.~(7) of \citet{SHInsuff}) for an illustrative 100 sq degree 10-bin tomographic weak lensing survey with a galaxy number density of $2.6/{\rm arcmin^2}$ per tomographic bin.  Greenish elements mark data points that are subject to non-Gaussian statistics. Redshifts increase to the lower right corner. Angular bins range from $0.5$ arcminutes to about 6 degrees. Within each redshift bin, the angular scale increases towards the lower right corner.  Left: without shape noise. Right: with shape noise. Top: $\xi_+$. Bottom: $\xi_-$. All colours are to scale. From the right we see that the non-Gaussianities are more prominent on large angular scales where the increasing number of galaxy pairs suppresses the shape noise.}
\label{comp}
\end{figure*}

In Sect.~\ref{Accuracy} we prove that our statistical model gives rise to a weak lensing likelihood, which passes four stringent tests that a Gaussian likelihood fails. The approximations discussed here hence represent the actual noise on weak lensing data more faithfully than a Gaussian approximation with an arbitrarily complicated covariance matrix. 
We therefore proceed to prove that our likelihood is not just a mere model, but a faithful representation of genuine weak lensing data.

\section{Accuracy of the derived likelihood}
\label{Accuracy}
\citet{SHInsuff} presented stringent tests that characterize non-Gaussian statistical behaviour, and any sound weak lensing likelihood should hence pass these tests. The methods derived in \citet{SHInsuff} test whether all pairwise combinations of data elements display the correct statistical behaviour under addition, division and multiplication, and whether the correct marginal distribution ensues for each data point. In combination, these tests reveal whether random variables have the correct mathematical behaviour as real data. \citet{SHInsuff} show that non-Gaussianities are present in the CFHTLenS data set and we here show that this is a generic feature of weak lensing data, and that our likelihood Eq.~(\ref{Like}) is able to reproduce these non-Gaussian features well.

Fig.~\ref{comp} depicts a trans-covariance matrix as first defined in \citet{SHInsuff}. A trans-covariance matrix has the same structure as a covariance matrix, but whereas a covariance matrix measures the covariance between two data points, a trans-covariance measures non-Gaussian correlations instead. The trans-covariance matrix of Gaussian data vanishes. The strength of trans-covariance matrices is that they hunt for non-Gaussian correlations not by computing cumulants but by computing distributions instead. As cumulants only carry incomplete and limited information on non-Gaussianities, the test via distributions is more robust and more sensitive. To be precise, \citet{SHInsuff} define in their Eqs.~(7), (10) and (12) three trans-covariance matrices, ${\sf S^+, S^\div, S^*}$, which test whether sums, ratios, and products of random variables follow the correct distribution. The elements of the trans-covariance matrices then depict the total deviation of the measured distribution from the distribution that should ensue. Trans-covariance matrices thereby react to whether the correct skewness is produced, the correct overall shape of the distribution, the correct outlier fraction and many more properties arising from a distribution's shape.

\begin{figure}
\includegraphics[width=0.48\textwidth]{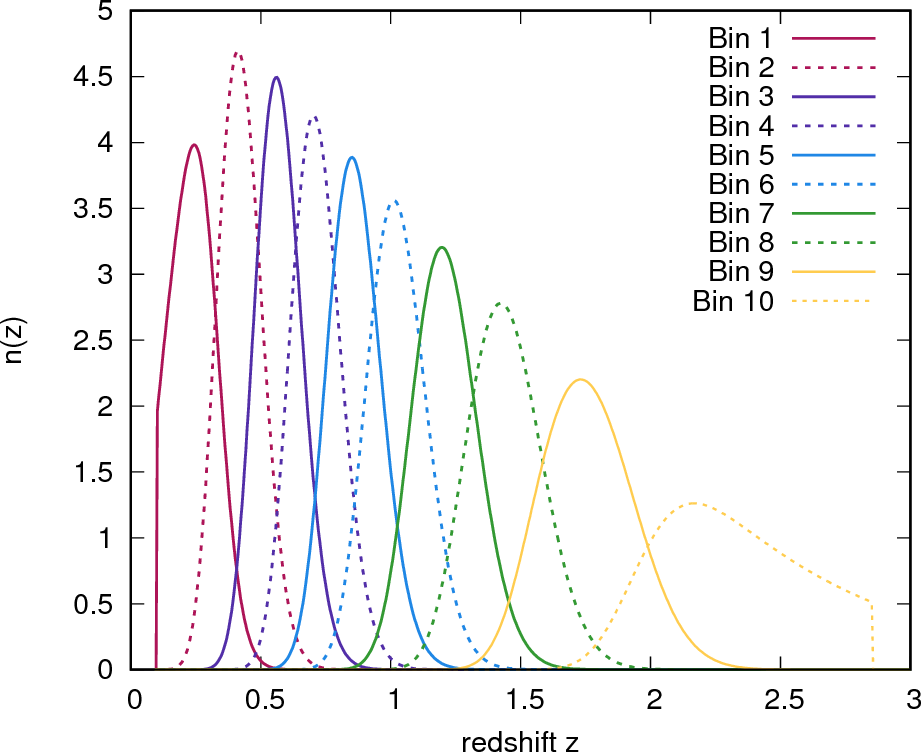} 
\caption{The ten tomographic redshift bins as chosen to model a typical weak lensing survey. The galaxy density is constant per bin with $2.6/{\rm arcmin^2}$.}
\label{LSSTBins}
\end{figure}

In Fig.~\ref{comp}, the trans-covariance matrix was estimated from a set of 930 simulated weak lensing data vectors. These 100 square degree field-of-view simulations are a significant advance on the 12.84 square degree `Clone-simulations' \citep{Clone1,Clone2,Clone3} used in \citet{SHInsuff}.  The non-Gaussianities here discussed therefore cannot be attributed to the known loss of power in the `Clone-simulations' for angular scales $\theta > 10\ {\rm arcmin}$ that resulted from the small simulation box size of $L = 147\ {\rm Mpc} \, {\rm h}^{-1}$.   With a simulation box size of $L = 505\ {\rm Mpc} \, {\rm h}^{-1}$, this finite box effect impacts the amplitude of the recovered SLICS weak lensing signal only on scales larger than $\theta > 100\ {\rm arcmin}$.  This now well understood effect is also corrected for in our analysis.

The trans-covariance matrix in Fig.~\ref{comp}, $\boS^+$, was produced for an illustrative 100 square degree weak lensing survey, assuming 10 tomographic bins with an equal number density of galaxies in each bin of $2.6/{\rm arcmin^2}$. The redshift bin modeling for this mock survey is depicted in Fig.~\ref{LSSTBins}, displaying 10 tomographic bins of equal number density, each  convolved with a Gaussian filter of width $\sigma = 0.02(1+z)$, and restricted to the redshift range [0.1 - 3.0]. The correlation functions $\xi_+$ and $\xi_-$ are used as summary statistics, with 32 angular bins, logarithmically spaced equidistantly between $0.5\ {\rm arcmin}$ and $400\ \rm{arcmin}$.

While the left panels of Fig.~\ref{comp} omit shape noise, the right-hand panels also contain a Gaussian shape noise of $\sigma_{\epsilon_i} = 0.29$ per ellipticity component. The colour bar indicates how strongly two data points couple in a non-Gaussian fashion, with green and yellow being data points that are most strongly subjected to non-Gaussian statistics. Data points indicated in blue behave approximately Gaussian under addition. The full trans-covariance matrices for division and multiplication also reveal the presence of non-Gaussianities, and are discussed in more detail in Section~\ref{sec:trans-tests}. 

\begin{figure}
\includegraphics[width=0.45\textwidth]{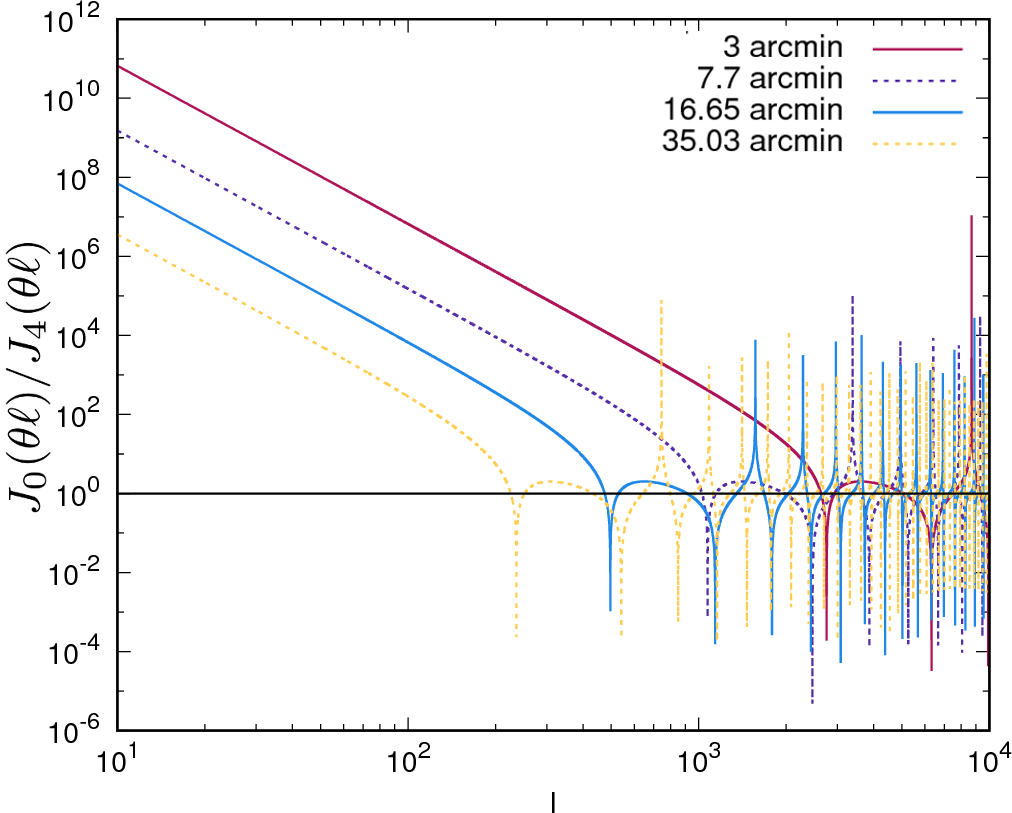} 
\caption{Ratio of the filter functions for $\xi_+$ and $\xi_-$. This ratio is equal to $J_0(\theta \ell)/J_4(\theta \ell)$, and for this plot the absolute value was taken. If the ratio is equal to unity, then $\xi_+$ and $\xi_-$ put the same weight on harmonic modes with a given $\ell$, but as can be seen, especially for low $\ell$, $\xi_+$ puts more weight on these modes. As fewer modes with low $\ell$ exist, this means the estimator $\xi_+$ will be more subject to non-Gaussianities.}
\label{filterplot}
\end{figure}

By comparing the top and bottom panels of Fig.~\ref{comp}, it is evident that the estimator $\xi_+$ is more subject to non-Gaussian correlations than $\xi_-$. This is caused by the different filter functions of $\xi_+$ and $\xi_-$. Fig.~\ref{filterplot} shows the ratio of the filter functions for $\xi_+$ and $\xi_-$ and illustrates that the filter of $\xi_+$ puts a larger weight on low $\ell$-modes. As these low $\ell$-modes have the smallest degrees of freedom $\nu$, they are most subject to skewness and thereby cause the striking non-Gaussianities in $\xi_+$. An exchange of the filter function to any other filter function commonly used in weak lensing \citep{Kilb,Cos17} or even an optimal linear compression thereof, e.g. via a MOPED-filter function \citep{Moped}, will change the non-Gaussianity of the compressed 2-point correlation function, and this is self-consistently modeled by our likelihood.

By comparing the left and right panels of Fig.~\ref{comp}, we see that the weak lensing likelihood is indeed strongly modular. This supports the structure of our likelihood Eq.~(\ref{Like}) which treats the randomness of cosmic structure formation, and the shape noise due to galaxies being non-spherical on separate levels. We furthermore see that shape noise leads to a strong Gaussianization of the data. This is expected, since the shape noise model that has been included in the SLICS simulations follows a Gaussian distribution.  The right-hand panels show that, in the presence of shape noise, the non-Gaussianity is only significant on large angular scales.  On these scales, which contain the most galaxy pairs, Gaussian shape noise is suppressed but the non-Gaussianities seen in the left-hand panels still remain. We therefore conclude that, for a fixed survey area, deeper surveys will suffer more significantly from these non-Gaussian statistics.

\begin{figure*}
\includegraphics[width=\textwidth]{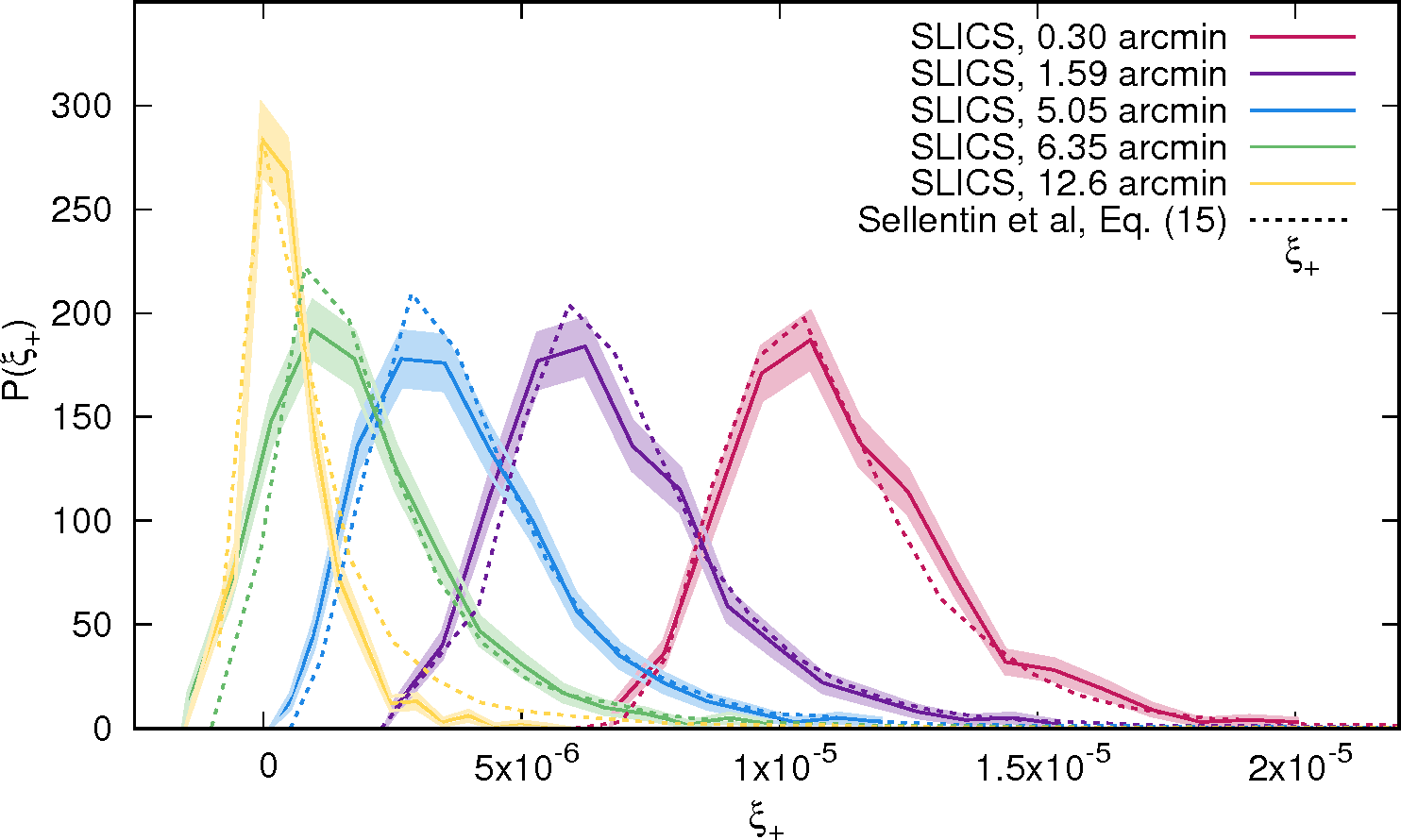} 
\caption{Non-Gaussian marginal distributions of individual weak lensing data points. Depicted is $\xi_+$ for different angles as given in the legend, for the first tomographic redshift bin. The solid curves are measured from the 100 square degrees SLICS with zero shape noise.  The dashed curves were produced from the non-Gaussian likelihood Eq.~(\ref{Like}) with the same settings as the simulations. This shows that our mathematically derived likelihood produces marginal distributions in agreement with the simulations. In other words, our assumption of the $a_{\ell m}$ being Gaussian-distributed is a good approximation. As can be seen, the density functions are strongly left-skewed which makes them peak below the mean. It is thereby very likely that one weak lensing data vector contains many data points which are lower than average.}
\label{Xips1}
\end{figure*}

Having thus supported the modular structure of our likelihood, we continue to show that our likelihood Eq.~(\ref{Like}) also succeeds in reproducing all marginal distributions of the simulated weak lensing data within the regime where the simulations are reliable (see Sect.~\ref{sec:SLIClims}). This can be seen in Fig.~\ref{Xips1}, which depicts the distribution of $\xi_+$ estimators for the first tomographic redshift bin in Fig.~\ref{LSSTBins}, assuming zero shape noise (i.e. $\sigma_\epsilon = 0$). The binning of the histograms which compare the distribution of $\xi_+$ measured from the SLICS simulations, with the distribution predicted by our likelihood function in Eq.~(\ref{Like}), is the same. For Fig.~\ref{Xips1}, the sky area of our likelihood was set to 100 square degrees, which is set by the size of the SLICS simulated sky patch.  We find the same level of agreement for all tomographic redshift bins, and for the case when shape noise is also included in the analysis.  We therefore conclude that our likelihood correctly reproduces the marginal distributions as produced via simulations of cosmic structure formation and weak lensing.

\begin{figure*}
\includegraphics[width=0.99\textwidth]{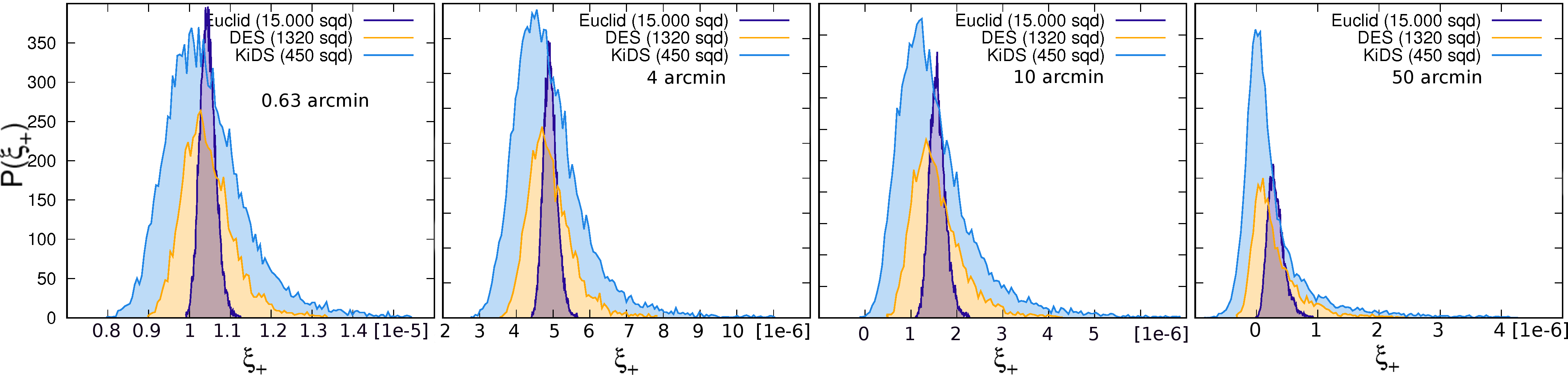} 
\caption{Increasing the sky coverage leads to a slow Gaussianization of the marginal distributions of $\xi_+(\theta)$. The distributions here shown were produced with our forward model Eq.~(\ref{Like}) and include the low $\ell$-modes that the SLICS simulations lack. The Gaussianization first sets in at small angular separations and proceeds to larger angular scales when the survey volume is increased. This is due to ergodicity, where the increasing volume average feigns an ensemble average. Even for a Euclid-like survey, noticeable non-Gaussianity will remain on angular scales above $\approx 50 {\rm \ arcmin}$. Shape noise has here not been added; the displayed distributions refer to redshift bin 1.}
\label{Gaussianization}
\end{figure*}

For the SLICS simulations, the number of possible histogram bins is limited by the available number of 930 simulations. In contrast, sampling from our likelihood is straight-forward, and Fig.~\ref{Gaussianization} therefore depicts a more highly resolved version of the shape noise free marginal distributions, adapted to a sky coverage of 450 square degrees (KiDS-450-like), 1320 square degrees (DES-Yr1-like) and 15000 square degrees (Euclid-like), where the mock survey properties are summarized in Table~\ref{Tab1}. Fig.~\ref{Gaussianization} reveals the strongly skewed nature of the marginal distributions. However for increasing sky coverage, a slow Gaussianization process sets in, such that the distributions for the Euclid survey area are more symmetric than those for the current KiDS and DES survey areas (but not fully symmetric). The Gaussianization arises from ergodicity: as the survey size increases, the areal average gets closer to the ensemble average. This is also why small angular scales Gaussianize first. The slow Gaussianization by increasing the survey volume however also implies that the skewness of weak lensing likelihoods is subject to the same constraints as cosmic variance: the finiteness of the maximally ever observable cosmic volume leads to a minimal asymmetry that the weak lensing distributions will maintain. 

\begin{table*}
	\centering
	\caption{Weak lensing survey properties which affect the likelihood.}
	\label{Tab1}
	\begin{tabular}{lccr} 
		\hline
		 Parameter & Value & Meaning & Influences \\
		\hline 
		$\bar{n}$ & $2.6/{\rm arcmin^2/tomogr.bin}$ & galaxy density  & Degrees of freedom via pixelization\\
		$g_{\rm eff}$ & calibrated & density of states  & Effect of mask geometry on degrees of freedom\\
		$A_{\rm sky}$ & 450/1320/15.000 deg$^2$ & survey area & Values adopted for KiDS/DES/Euclid-like surveys\\
		$f_{\rm sky}$ & $A_{\rm sky}[{\rm sterad}]/4\pi$ & sky fraction & Degrees of freedom by setting density of $\ell$ modes\\
		$\sigma_{\epsilon_i}$ & 0.29 & shape noise std. dev. & Reduction of overall precision due to shape noise\\
		\hline
	\end{tabular}
\end{table*}

\subsection{Trans-covariance tests}
\label{sec:trans-tests}
\begin{figure*}
\includegraphics[width=\textwidth]{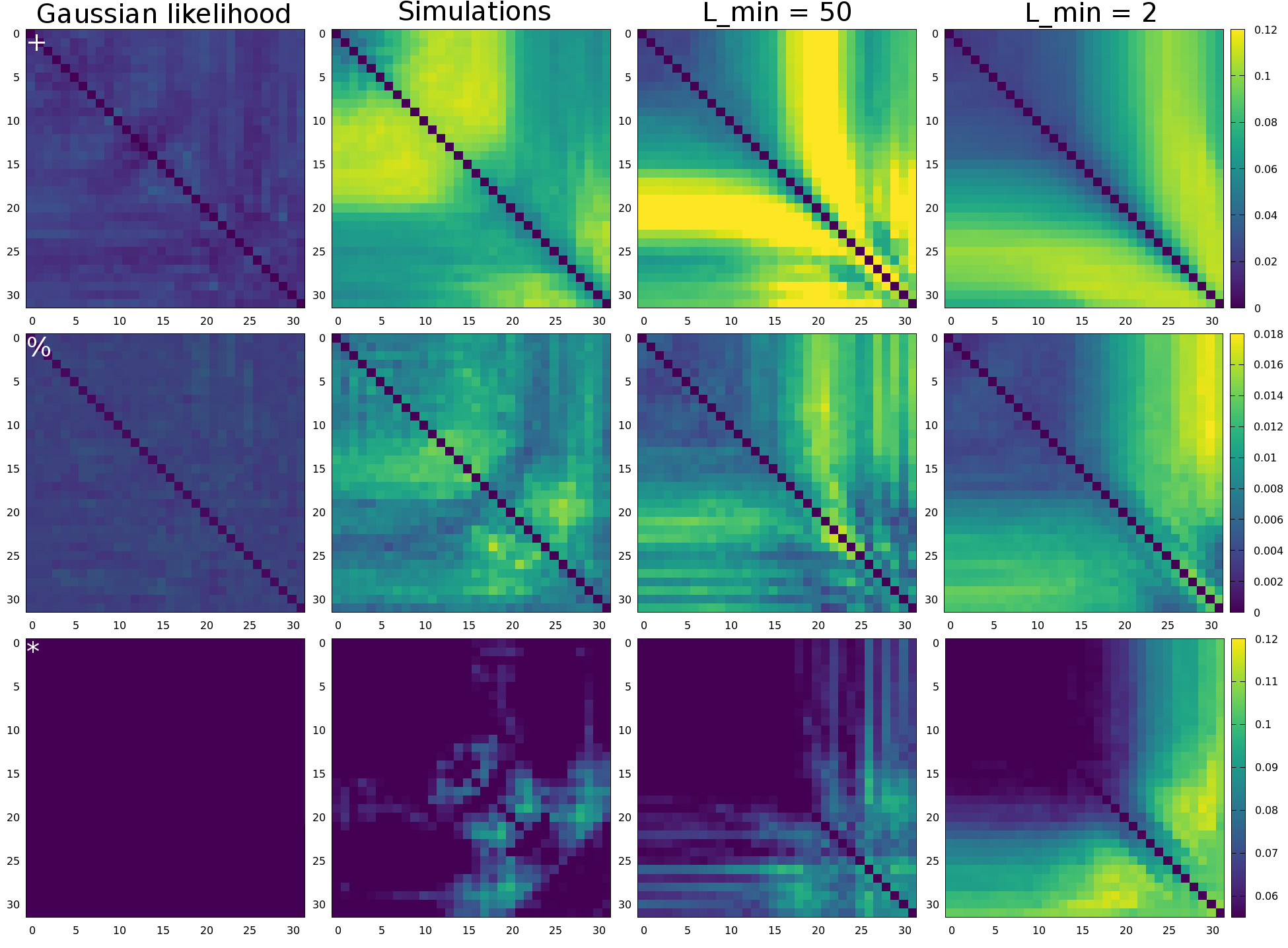} 
\caption{Trans-covariance matrices, a highly sensitive and cosmology-independent test of non-Gaussianity \citep[see][]{SHInsuff}). The correct weak lensing likelihood should -- apart from residual inaccuracies in the numerical simulations -- reproduce the patterns seen in the second row. Rows: Addition, division, multiplication. First column: vanishing trans-covariance matrices of the currently standard Gaussian likelihood. Second column: trans-covariance matrices measured from the SLICS simulations, whose non-zero elements indicate whether the sums, ratios and products follow the correct distribution. By definition, the Gaussian likelihood from the first row cannot produce the patterns of the second row. Third column: trans-covariance matrices derived from our likelihood Eq.~(\ref{Like}), where we approximately mimic the loss of power in the simulations by turning off the low $\ell$-modes. The up to $\sim 10\%$ deviation of the simulated SLICS data from the input cosmological model due to resolution effects is not corrected for, contributing to the discrepancies between the second and the third column. Nevertheless, our likelihood is a significant improvement in recovering the simulated trans-covariance matrices, in comparison to the Gaussian likelihood. Fourth column: Turning on the low $\ell$-modes that are lost in simulations enhances the non-Gaussianity because the low multipoles have the lowest degrees of freedom. The simulations therefore underestimate the skewness of weak lensing likelihoods. The plots refer to $\xi_+$ of the first redshift bin and are representative of the other redshifts. Here, no shape noise was added. The agreement between simulations and our likelihood Eq.~(\ref{Like}) improves with shape noise, see Fig.~\ref{TranscWSN}.}
\label{PassTest}
\end{figure*}

To test whether our likelihood generates data that also have the correct non-Gaussian behaviour when convolved in mathematical functions, we conduct the trans-covariance tests of \citet{SHInsuff} on them. The results are displayed in Fig.~\ref{PassTest}, and are to be interpreted as follows: If a likelihood fails the trans-covariance tests, it means this likelihood assesses correlations in the data incorrectly. Such a likelihood then does not correctly account for mutual inter-dependencies between different data points. In analyses of the large-scale structure, this is especially problematic, as our signals are precisely correlations between different data points. In general, a likelihood that fails the trans-covariance tests is incorrectly shaped, which will lead to biases when inferring physical parameters. It will also lead to biases when inferring which physical model is preferred out of a set of multiple competing models.

\begin{figure}
\includegraphics[width=0.49\textwidth]{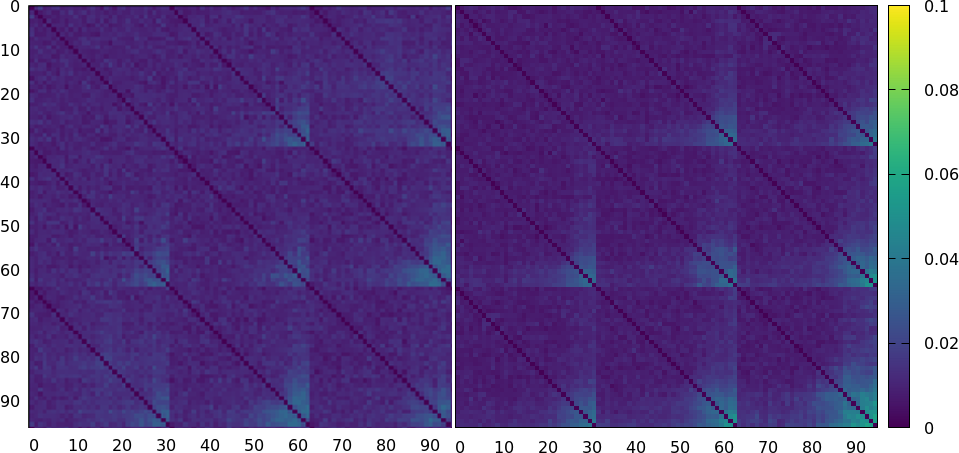} 
\caption{Transcovariance matrix with shape noise; displaying the first 3 redshift bins from Fig.~\ref{comp}. Left: SLICS simulations. Right: our likelihood Eq.~(\ref{Like}). The right-hand panel also includes the low $\ell$-modes that the SLICS simulations lack. }
\label{TranscWSN}
\end{figure}

The tests in Fig.~\ref{PassTest} demonstrate that the samples drawn from our likelihood Eq.~(\ref{Like}) represent the statistical behaviour of weak lensing data more faithfully than the Gaussian likelihood. Samples from our likelihood add correctly, multiply correctly and also have the correct ratios. As all mathematical functions can be represented by a concatenation of these operations, we therefore conclude that our likelihood from Eq.~(\ref{Like}) represents the true statistical behaviour of weak lensing data to a high degree of accuracy.

In contrast, the first column of Fig.~\ref{PassTest} illustrates that a Gaussian likelihood fails the trans-covariance tests, irrespective of whether the covariance matrix was computed by any of the methods presented in \citet{Lacasa1,Lacasa2,Covs,KiDS,DES,SH15,SH17}. This means weak lensing data are non-Gaussianly distributed and assuming a Gaussian likelihood therefore poses an unnecessary limitation to the quality of weak lensing analyses. This holds true for even arbitrarily precise covariance matrices.

\subsection{Known limitations of the SLICS simulations}
\label{sec:SLIClims}
In order to facilitate the comparison between our likelihood and the synthetic weak lensing data vectors from the SLICS simulations, we compile in Table~\ref{Tab2} the known limitations of the simulations. The most influential problem with the simulations is that the finite simulated volume cannot support large-scale modes, and hence power is lost with low $\ell$-modes.  When comparing our likelihood to the SLICS simulations, we mimic this effect in our likelihood by switching off the low $\ell$-modes.  This is an approximation as the low-$\ell$ power in the simulations decays smoothly, in contrast to the sharp cutoff that we implement.  As such we should not expect perfect agreement between the high $\theta$ marginal distributions shown in Fig.~\ref{Xips1}.  Nor should we expect the trans-covariance matrices measured from the SLICS simulations (second column of Fig.~\ref{PassTest}) to perfectly agree with the trans-covariance matrices measured with the inclusion of this $\ell$-mode cut in our likelihood (third column of Fig.~\ref{PassTest}).  

Even though the low $\ell$-modes are missing in the simulations, they are of course present in real data, and we hence display the trans-covariance matrices for all $\ell$-modes down to $\ell = 2$ in the last column of Fig.~\ref{PassTest}. As expected, the non-Gaussianity in the weak lensing data increases when these low $\ell$-modes are included. The low $\ell$-modes also dominate the skewness of the marginal distributions, which is why Fig.~\ref{Xips1} (where we switched off the low multipoles to mimic the simulations) displays less skewed marginal distributions than Fig.~\ref{Gaussianization} where we plot marginal distributions after including the low $\ell$-modes that the simulations exclude. 

A further more difficult to mimic discrepancy between the simulations and our likelihood is that the resolution effects in the simulations produce $\xi_+$ and $\xi_-$ realizations which deviate by up to $\sim 10\%$ from the theoretical prediction of $\xi_+$ and $\xi_-$ from the input cosmology. This is also understood and is described in more detail in \citet{SLICS}.  We measure the mean scale dependent deviation between the simulated and theoretical prediction for $\xi_+$ and $\xi_-$.  We then add this mean value to each SLICS realisation when comparing results in Fig.~\ref{Xips1}.  This correction translates the histograms, but does not change their shape. 

As this correction may affect the trans-covariance matrices in an opaque way (because various histograms are then shifted with respect to each other), this calibration is not included in Fig.~\ref{PassTest}. The remaining expected discrepancy between the simulations and our likelihood (which truly samples from the SLICS input cosmology), is therefore partially responsible for the relatively small differences in the trans-covariance matrices that compare our likelihood to the simulations. 

Overall, it can be seen that our likelihood is more reliable in reproducing the trans-covariance matrices and the marginal distributions, than a Gaussian likelihood. The trans-covariances in Fig.~\ref{PassTest} refer to a shape noise free case, i.e. they only test the non-Gaussianity of the likelihood arising from noise in the large-scale structure that feeds through to the 2-point function. Fig.~\ref{TranscWSN} demonstrates that when adding shape noise, the quantitative agreement between our likelihood and the simulations increases further. This is because shape noise dominates on the smallest angular scales. Our Eq.~(\ref{Like}) is therefore also likely to be robust with respect to non-linearities in structure formation, because these will primarily affect the likelihood arising from noise in the large-scale structures on the smallest of angular scales, where shape noise is the dominant uncertainty.

Even though the limitations of the SLICS simulations are understood and can be mimicked, a certain grey-zone arises, where it is not completely clear how the output of the simulations should be interpreted. This affects the comparison with our likelihood. For future improvements of the likelihood here presented, we hence target to become independent of the calibration on the simulations, and compute the degrees of freedom in a fully independent manner.

\begin{table*}
	\centering
	\caption{Limitations of the SLICS simulations which affect the comparison with the likelihood here derived.}
	\label{Tab2}
	\begin{tabular}{lccr} 
		\hline
		 Parameter & Value & Meaning & Causes \\
		\hline 
		$k_{\rm Nyq}$ & $19 h/{\rm Mpc}$ & Fourier Nyquist frequency & Simulations lack scatter from high $k$ modes.\\
		$\ell_{\rm Nyq}$ & $1.3 \cdot 10^5$ & harmonic Nyquist frequency & Simulations lack scatter from high $\ell$.\\
		$\ell_{\rm min}$ & $\approx 20$ & smallest $\ell$ resolved & Simulations lack skewness from low $\ell$ modes.\\
        $A_{\rm sky}$ & 100 deg$^2$  & simulated survey area & Sampling range by cutting of high $\ell$ modes\\
		$m_{\rm res}$ & $2.88 \cdot 10^9 M_\odot/h$ & Mass resolution & Scale-dependent deviation of $\xi_+,\xi_-$ from the input cosmology.\\
 $\Delta\xi_+$ & $\approx 10\%$ & typical discrepancies & Angular dependent uncertainty of simulated $\xi_+$.\\
		\hline
	\end{tabular}
\end{table*}

\section{Implication for weak lensing biases}
Given that our likelihood for weak lensing 2-point functions successfully passed 4 tests that a Gaussian likelihood fails, we are now in a position to use a fully probabilistic framework to determine how parameter biases in weak lensing will arise when using a Gaussian likelihood.

\begin{figure*}
\includegraphics[width=0.99\textwidth]{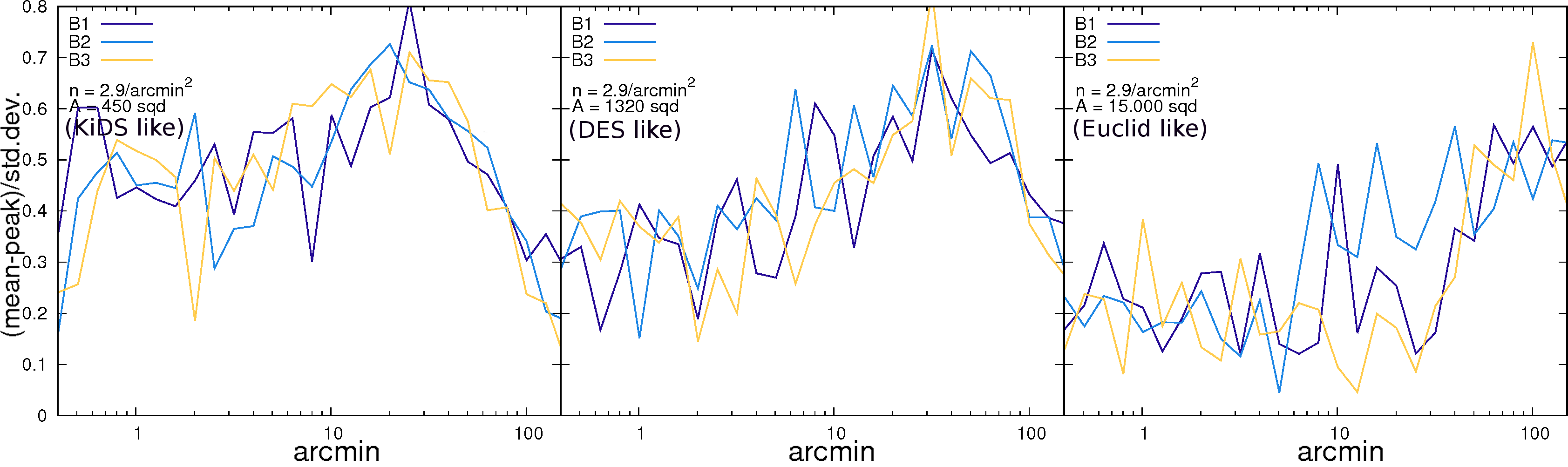} 
\caption{Bias of the shape-noise free weak lensing correlation functions, divided by the standard deviation per data point. This is the ratio of biases arising from the large-scale structure, to variance caused by the large-scale structure. The legend B1, B2, B3 refer to redshift bins 1,2,3 from Fig.~\ref{LSSTBins}. As can be seen, each data point is consistently biased low by around half a standard deviation. This ratio between the bias and the standard deviation remains remarkably constant when increasing the sky coverage from 450 square degrees (KiDS-like) to 15.000 square degrees (Euclid-like). However, Gaussianization sets in very slowly, see also Fig.~\ref{Gaussianization}. The noisiness of the curves arises because all samples of a histogram contribute to mean and standard deviation, while only the samples in the highest bins influence the peak. The peak position is hence noisier than the mean, and this noise has purposefully been left in the plots to give a visual impression of the uncertainty. The biases are relatively universal for different redshift bins because the redshift binning determines the shear-$\hat{C}_\ell$, but the bias is dominated by the degrees of freedom $\nu$, which react to survey size, rather than survey depth and binning. Cross-bins are therefore similarly biased as the intra-bin correlations here shown.}
\label{Biases}
\end{figure*}

\begin{figure*}
\includegraphics[width=0.99\textwidth]{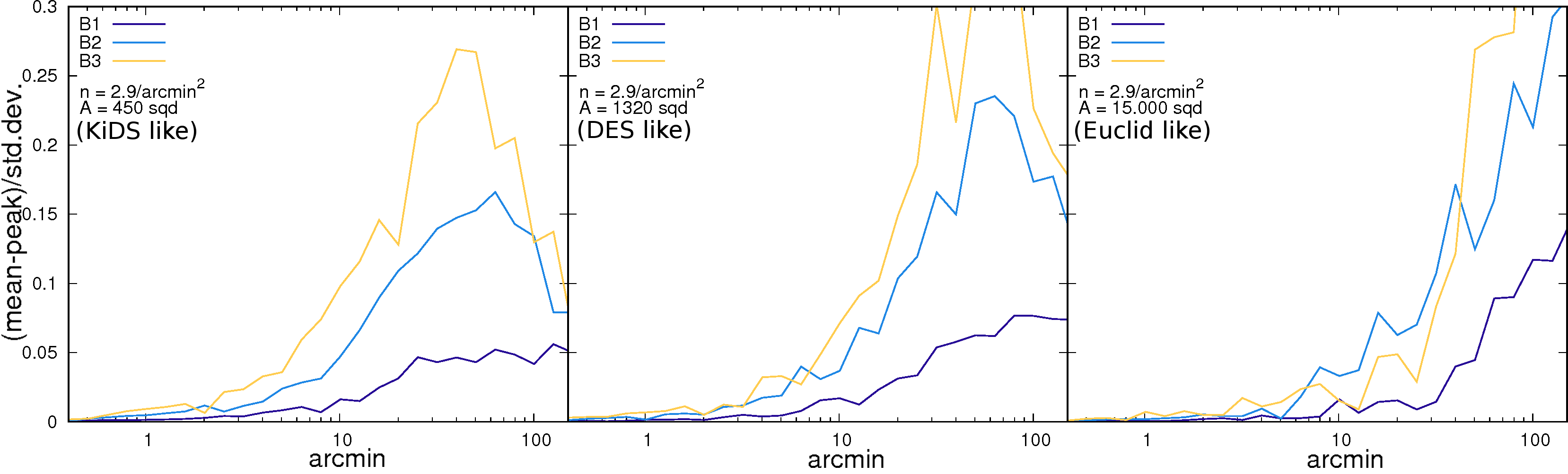} 
\caption{Like Fig.~\ref{Biases}, but now with shape noise added: this does not remove the bias which arises from cosmic variance of the 2-point function, but it reduces the ratio of the bias to the now increased total standard deviation. On the smallest angular scales, the bias is negligible in comparison to the total noise, yet for scales larger than 20 arcmin. The data of a KiDS-like survey can be biased by about 10\%-20\% of a standard deviation, and up to 25\%-30\% for surveys with DES-like and Euclid-like sky coverage on the same angular scales. For each sky coverage in the Figures \ref{Gaussianization},\ref{Biases} and \ref{BiasesWSN}, 26.000 samples of a tomographic survey with the first three bins from Fig.~\ref{LSSTBins} were produced. Per survey, this required 200 single-core CPU hours, illustrating that our likelihood is sufficiently fast to allow for improvements.}
\label{BiasesWSN}
\end{figure*}

The marginal distributions in Fig.~\ref{Xips1} and also those in Fig.~\ref{Gaussianization} illustrate that the weak lensing likelihood is left-skewed. This means it will most often produce data vectors whose lensing amplitude is lower than that of the input cosmology. These biases are quantified in more detail in Fig.~\ref{Biases} and Fig.~\ref{BiasesWSN}: these figures illustrate that the value of each data point is biased low, such that the biases occur already before parameter inference is conducted. 

This bias is independent of any astrophysical or data-related systematics and cannot be addressed by, for example, improvements in shape measurement technology, or by marginalising over nuisance parameters in a standard Gaussian likelihood analysis. This bias neither indicates any modifications of gravity or other deviations from $\Lambda$CDM: \emph{each} input cosmology will always generate weak lensing data that are most likely below their mean. The bias is caused by our analysis technique computing a 2-point function, and it is in this sense `self-made' but unavoidable. 

The left-skewness arises because the shear power spectra $\hat{C}^{\mu\nu}_\ell$ are Gamma distributed. The Gamma distribution is a non-symmetric distribution and peaks below its mean. Its mean ($\mu$), peak ($p$) and variance ($v$) are given by
\begin{equation}
\begin{aligned}
& \mu  = C_\ell \\
& p  = C_\ell\left(1-\frac{2}{\nu} \right) \ \ {\rm for\ \ } \nu > 2\\
& v  = \frac{2C_\ell^2}{\nu}.\\
\label{peak}
\end{aligned}
\end{equation}
From Eq.~(\ref{peak}) we see that as $\nu$ increases, the expectation value approaches the peak value.  For an infinitely precise measurement, the maximum likelihood estimator is therefore unbiased. For finite $\nu$ however, the bias $B = \mu - p$ is 
\begin{equation}
 B = -2C_\ell/\nu.
\end{equation}
The fact that the different $C_\ell$ are all continually biased low then feeds through to the real-space correlation functions, such that these are also biased low. The left-skewness implies that the maximum-likelihood parameters will not coincide with the mean parameters. Only for large degrees of freedom $\nu$ will the Central Limit Theorem kick in, and the skewed Gamma distribution then tends to its Gaussian limit, given by
\begin{equation}
 -2 \log \mathcal{L}(C_\ell | \hat{C}_\ell  ) = \frac{\nu}{2}\left( \frac{C_\ell - \hat{C}_\ell}{\hat{C}_\ell} \right)^2.
\end{equation}
The distribution of real-space correlation functions will then Gaussianize accordingly, and in this Gaussian limit, the peak of the likelihood and the mean coincide. As the degrees of freedom increase linearly with $f_{\rm sky}$, the likelihood will only linearly tend back towards a Gaussian as the survey area grows. In Figs.~\ref{Gaussianization} and \ref{Biases} we see that not even a Euclid-like survey area reaches this Gaussian limit fully.  Structure formation by itself leads to each data point of $\xi_+$ being biased low by up to half a standard deviation, in the absence of shape noise. 

Adding shape noise does not remedy this situation as shape noise does not decrease the absolute value of the bias. It does however decrease the ratio of the bias to the total statistical error by increasing the uncertainty on the measurement.

The discussion until now has been independent of a theoretically motivated parameterization to explain the data. In fact, the input cosmology was so far only needed to generate data, but was kept fixed throughout the entire paper. Taking the remaining step of translating the biases from the data onto biases of physical parameters, is now trivial: As the weak lensing amplitude scales with $S_8 = \sigma_8 \sqrt{\Omega_{\rm m}}$, the low weak lensing amplitude directly translates into a maximum likelihood estimator for $S_8$ being biased low, and the smaller the degrees of freedom, the larger this bias. 

\begin{figure*}
\includegraphics[width=\textwidth]{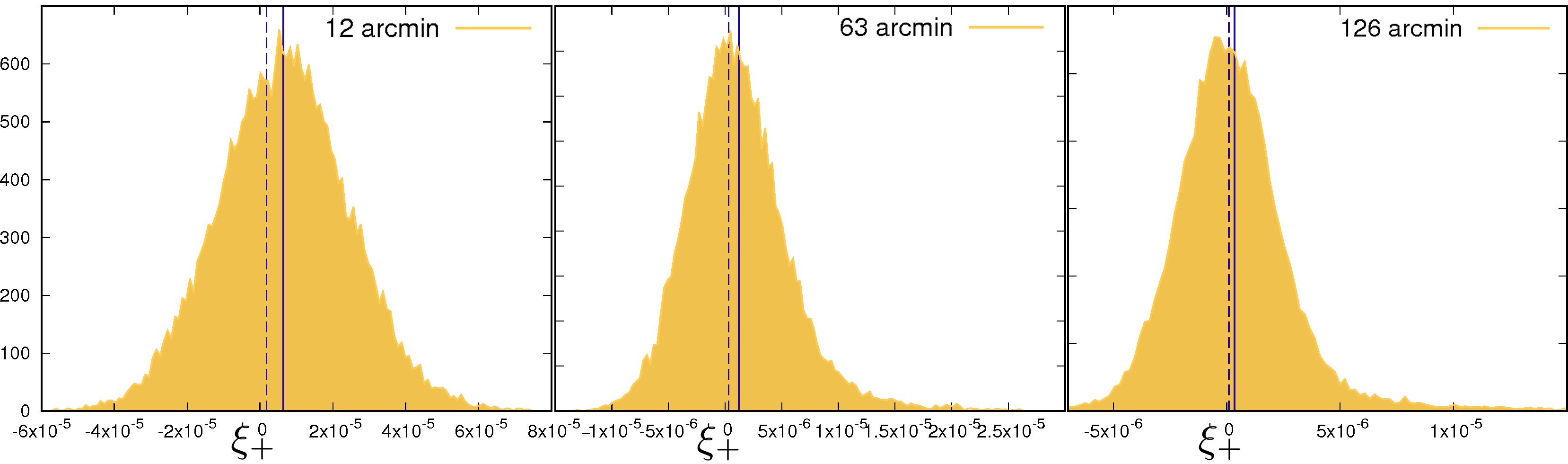} 
\caption{Marginal distributions of $\xi_+(\theta)$ for the third redshift bin in Fig.~\ref{LSSTBins} assuming a 450 square degree survey with shape-noise.  The skewness of the likelihood translates into a bias on $\sigma_8$ that depends on angular scale. The solid blue line indicates the predicted value of $\xi_+(\theta)$ for the SLICS input cosmology which has $\sigma_8 = 0.826$.  The dashed line indicates the predicted value of $\xi_+(\theta)$ for the same set of cosmological parameters, but with $\sigma_8$ halved to $\sigma_8=0.413$. Left: on small angular scales, with $\theta = 12\, {\rm arcmin}$, ergodicity removes the bias, such that the likelihood peaks at the input $\sigma_8$. Middle: on intermediate scales, with $\theta = 63\, {\rm arcmin}$, the skewness of the likelihood can shift the peak from the input $\sigma_8$ to half its value. Right: on the largest angular scales in comparison to the survey footprint, with $\theta = 126\, {\rm arcmin}$, the variance increases sufficiently to reduce the difference between the mean and peak of the distribution in comparison to the scatter.}
\label{Sig8}
\end{figure*}

The total bias on $\sigma_8$ is an agglomeration of the biases contributed by each data point, and the bias of each data point depends mainly on angular scale, the galaxy density of the survey, and the area of the survey -- it depends relatively weakly on the shear power spectra themselves. Fig.~\ref{Sig8} illustrates that the left-skewness of the weak lensing likelihood indeed has the correct order of magnitude to explain the low $\sigma_8$ as found in weak lensing studies. It has however to be cautioned, that the arising bias is stochastic, i.e. it will depend on the realization of the data vector drawn. The meaning of this is discussed in the conclusions, together with an outlook of how weak lensing data can be manipulated to reduce the impact of this bias on physical parameters.

\section{The road ahead}
Data analysis can only lead to unbiased constraints on a physical theory if the employed likelihood is correct~\citep[e.g.][]{Trotta,MardiaKentBibby,AndersonTW,CramerH,Jeffreys,SunBerger}. At the same time, it is broadly recognized that weak lensing is subject to uncertainties which should ideally be fully and consistently included in the analysis, but the current standard approach via a Gaussian likelihood does not allow this. One example for this is the addition of a shape noise covariance matrix onto the covariance matrix of cosmic structure formation: the addition of covariance matrices is only correct, if the noise distributions of both elements are Gaussian. Another challenge that is currently not handled fully and consistently are intrinsic alignments: these produce correlated shapes between galaxies, and thereby modify both the noise and signal at the same time. Only including them in the signal is insufficient.

In order to enable a consistent treatment of such weak lensing uncertainties, we have dropped the assumption of a Gaussian likelihood, and updated to a modular distribution based likelihood. 
Apart from being mathematically more principled, updating to a modular likelihood offers several major conceptual advantages.

Realistic shape noise is not well modeled by a Gaussian distribution \citep[e.g.][]{Shape3}. The current standard Gaussian likelihood however adds the standard deviation of shape noise to the total covariance, i.e.~it treats shape noise as if it were Gaussian distributed. Using our likelihood Eq.~(\ref{Like}), we see that the addition of covariances is insufficient as the third line of Eq.~(\ref{Like}) reveals that shape noise leads in reality to a convolution.  This arises because the sum $s = u+v$ of two independently distributed summands with $u \sim \calP_{u}(u)$ and $v \sim \calP_{v}(v)$, is distributed according to the \emph{convolution} of the two individual distributions 
\begin{equation}
s \sim \calP_s(s) = \int \calP_v(s - u)  \calP_{u}(u) \mathd u.
\end{equation}
This convolution leads to a smooth deformation of $\calP_s(s)$, such that the sum is distributed according to a genuinely new distribution. A likelihood that is a priori restricted to a Gaussian shape, cannot
include the emergence of such a genuinely new distribution. The current standard approach hence cannot account for this effect, whereas our likelihood correctly includes it. This remains true if we update to a more realistic non-Gaussian distribution of shape noise.

A further advantage of our modular likelihood is that redshift uncertainties can be included as well: in a Gaussian likelihood redshift uncertainties can only imperfectly be accounted for by marginalizing over nuisance parameters. Here however, uncertainties on $n_\mu(z)$ can be included in the second line of Eq.~(\ref{Like}).

In this manner, we see that the mathematical structure of our likelihood Eq.~(\ref{Like}) is more flexible than a Gaussian likelihood, enabling us to greatly refine weak lensing analyses. We therefore expect the mathematical framework here presented to become the core likelihood to future sequential refinements.  A current conceptual disadvantage, that we wish to improve upon, is the dependency on simulations to calibrate the number of degrees of freedom $\nu$ in the case of a masked survey geometry through the factor $g_{\rm eff}(\ell)$ in Eq.~(\ref{eqn:geff}).  Future refinements will address this dependence.

\section{Conclusions}
\label{Conclusions}
The tension between weak lensing constraints of $S_8 = \sigma_8/\sqrt{\Omega_{\rm m}}$ and CMB measurements from the Planck satellite motivates the question of whether weak lensing results are inherently biased low.  In this paper we have demonstrated that the answer to this question is yes. 
The amplitude of weak lensing 2-point functions is biased low already prior to any parameter inference -- then using a Gaussian likelihood approximation centres on this preferentially low amplitude and hence biases weak lensing $S_8$ constraints low by up to 30 percent of the weak lensing errors per data point. The framework derived in this paper allows us to discuss this question in great detail, and we have seen that these biases arise even if the cosmological theory is correct, and the data are sound.

It had previously been demonstrated that the likelihood of weak lensing 2-point functions is skewed \citep{SHInsuff}, and here the core elements of the skewed distribution have been derived. Mathematically it is a sum over weighted Gamma distributions, where the weights are given by the angular filter functions. The skewness of a likelihood for 2-point functions is already known from analyses of the CMB, where primarily the lowest multipoles are subject to asymmetric distributions \citep{Hamimeche1,Hamimeche2,BJK1,BJK2}. The exact same mechanism applies to weak lensing 2-point functions, only that their distributions are more asymmetric than those of a CMB data set with the same sky area. The enhanced asymmetry is caused by galaxies being discretely spaced tracer particles instead of a smooth random field.

The core elements of the skewed weak lensing likelihood are given in Eq.~(\ref{Like}), and currently include cosmic variance from the large-scale structure, and Gaussian shape noise from the intrinsic ellipticity dispersion of galaxies and the associated shape measurement uncertainty. A generalization to also include redshift misestimates, intrinsic alignments, mask geometry, additive and multiplicative biases, as well as selection effects during shear measurements, is intended as future work. A further intended refinement is the abstraction from simulations: even though we were here able to correct for known imperfections of simulations to an adequate precision for this paper, prior to applying our likelihood to data, we will pursue a mathematical study of how the degrees of freedom and the mask geometry interplay. Becoming independent of the simulations is certainly an important goal.

From the perspective of a statistician, having derived the non-Gaussian likelihood is a milestone which allows us to render data analysis more realistically. However, our result is somewhat confusing from the perspective of a theoretical physicist: dealing with an asymmetric likelihood means that the most likely parameters will not coincide with the parameters that are preferred on average. From three different perspectives, we henceforth discuss the meaning of biases arising in this manner.

The term `bias’ is typically used to describe a deterministically arising bias.  One example of a deterministic bias would be to neglect intrinsic alignments in the data analysis, despite knowing of their importance, and thereby forcing the likelihood to peak in an incorrect place. Accounting for the missed intrinsic alignments then corrects the analysis and thereby removes this bias.

The situation highlighted in this paper however deals with a different type of bias: the skewed weak lensing likelihood gives rise to \emph{stochastic} biases, rather than deterministic ones. A stochastic bias arises when the mean of a data set does not coincide with the most likely data set. It is therefore the data set itself which displays statistically biased behaviour. For a stochastically arising bias, there cannot exist a single satisfactory recipe to `remove' it: the success of such a recipe would always depend on the realization of the data vector drawn. For example, with the left-skewed likelihood here at hand, the maximum-likelihood estimate will be biased low, and one could hence be tempted to `correct' for this by adding in the average distance between maximum-likelihood and mean. This would move the data upwards through the distribution. If however, by chance, a dataset is drawn which already falls above the mean, then this upward-shifting `correction' recipe actually moves the data even further away from the mean, whereby the recipe fails. 

Such stochastically arising biases can only be treated and understood in a stochastic manner. In fact, for a strict Bayesian, the notion of a `bias' does not even exist: if the entire analysis is conducted conditional on the single data vector available, and a repetition of the experiment is impossible, then the notion of a `mean' is nonsensical, and thereby the notion of a bias. Instead, a Bayesian would not perceive the skewed likelihood as a problem and rather analyse it jointly with a prior. The maximum-a-posteriori parameters would then be interpreted as the `most likely' estimate of the underlying parameters which the Universe actually obeys. Importantly however, the resulting inference would then be interpreted as \emph{conditional} on the priors and the likelihood, and this conditionality is easily accepted by a Bayesian (but not by a theoretical physicist).

To a Frequentist, the skewness of the likelihood would be highly suspect. It means that the event that is most likely to occur, is not the event that occurs on average. The Frequentist would then prefer the mean as the best representation of the underlying model, and this will systematically deviate from the result of a Bayesian via maximum a-posteriori likelihood estimation. To opt out of this ambiguous situation, the Frequentist would hence repeat the measurements and average the data, whereby the likelihood of the average data set begins to Gaussianize. Thereby the Frequentist can asymptotically evade the problem of the stochastic biasing, but for cosmology this route is only viable within the limits of ergodicity and cosmic variance.

To the theoretical physicist, the skewed likelihood poses a conundrum: coming from a Lagrangian theory of the Universe and believing in the concept of there existing a \emph{unique} set of `true' parameters which have to be found, it is now highly undesirable that two \emph{ambiguous} concepts exist of what might represent the `true' parameters best: are the true parameters those that describe the Universe on average, or those that describe the Universe that is most likely? A theoretical physicist would typically also dislike the Bayesian conditionality interpretation or any heuristic `correction scheme', and rather prefer an unconditional statement on what the true parameters are (i.e. the theoretician would hope that the statistical inference leaves no traces when constraining physics). 

Having thus discussed the perception of the skewed likelihood from the three perspectives of a Bayesian, a Frequentist and a theoretical physicist, we here conclude that an interesting approach to deal with these stochastically arising biases is the following way:

If we wish to return to a unique `best fit' solution of parameters, then a symmetric likelihood is needed. It does not necessarily need to be Gaussian, other symmetric likelihoods such as the t-distribution presented in \citet{SH15,SH17} also lead to a unique best fit, while also correctly including noise in covariance matrices. For weak lensing 2-point functions, returning to a symmetric likelihood can be achieved in a three-step manner. First, the correlation functions $\xi_+$ and $\xi_-$ are to be measured in real space. These will however be biased in a complicated way, as all real space data points include contributions from low $\ell$-modes, which are the dominant source of the likelihood's asymmetry. Hence, one would include a potential second step of filtering the measured $\xi_+$ and $\xi_-$ through COSEBIS \citep{Cos10,Cos17}, in order to remove potential $B$-mode contributions from the signal. In harmonic space, the bias can then be studied and potentially be reduced by excluding the low multipoles. 

The last step of this procedure can lead to a symmetrization of the likelihood without loosing too much data. A coarse binning in real-space could of course also be done, however this would have the predominant aim to suppress asymmetry from low $\ell$-modes, and hence more constraining power of the data might be lost than by cleaning the data set in Fourier space. The resulting cleaned data set can then either be analyzed with the likelihood here presented, or depending on the success of the symmetrization, also with its Gaussian approximation. In total, making the best-fitting parameters unique in the manner described here, constitutes however a major re-analysis of weak lensing data sets. It is therefore scheduled for future KiDS cosmic shear analyses. Due to non-Gaussianities being a general feature of weak lensing, they are of equal importance for future DES \citep{Troxel} and Hyper Suprime-Cam \citep{HSC} studies of weak lensing.

The results derived in this paper, especially the numbers given for the biases, depend on the employed filter functions (here $J_0$ and $J_4$), which can however be quickly exchanged. The results also depend on the assumed sky coverage and the galaxy number density of the respective survey. The \emph{absolute} value of the biases can only be decreased by increasing the sky coverage and the number density of the surveys. The ratio of the biases to the total uncertainty displays a less monotonic behaviour: as the surveys become more precise and reduce shape noise and increase the sky coverage, the biases become ever more important in comparison to the total error bar. Depending on redshift, biases of up to 30\% of the standard deviation per data point are possible. This translates into scale-dependent biases on $\sigma_8$ that is typically biased low.
In total, we can maintain, that any sound cosmology produces weak lensing data whose amplitude is biased low, and this neither indicates a flaw in the data, nor a flaw in the cosmological theory. Strategies to address these biases could either be a probabilistic propagation via the likelihood here presented, or a restructuring of the analysis strategy to evade biased scales.

\section{Acknowledgements}
It is a pleasure to acknowledge scientific discussions with Ruth Durrer, Andrew Jaffe, Alan Heavens, Justin Alsing and Matthias Bartelmann. Papers by Antony Lewis and Samira Hamimeche were particularly influential during this research. ES is supported by the Swiss National Science Foundation. JHD is supported by the European Commission under a Marie-Sklodowska-Curie European Fellowship (EU project 656869). CH acknowledges support from the European Research Council under grant number 647112. Computations for the N-body simulations were performed on the TCS supercomputer at the SciNet HPC Consortium. 
SciNet is funded by: the Canada Foundation for Innovation under the auspices of Compute Canada; 
the Government of Ontario; Ontario Research Fund - Research Excellence; and the University of Toronto.

\bibliographystyle{mn2e}
\bibliography{TDist}

\label{lastpage} 
\bsp 
\end{document}